\documentstyle[aps,preprint,tighten,epsf,floats]{revtex}
\begin{document}
\draft
\hyphenation{wave-guide}
\title{Reflectance Fluctuations in an Absorbing Random Waveguide}

\author{T.~Sh.~Misirpashaev$^{a,b}$ and C.~W.~J.~Beenakker$^a$}
\address{$^a$Instituut-Lorentz, University of Leiden,
P.O. Box 9506, 2300 RA Leiden, The Netherlands\\
$^b$Landau Institute for Theoretical Physics, 2 Kosygin Street,
Moscow 117334, Russia}
\maketitle

\begin{abstract}

We study the statistics of the reflectance
(the ratio of reflected and incident intensities)
of an $N$-mode disordered waveguide
with weak absorption $\gamma$ per mean free path. Two distinct
regimes are identified. The regime $\gamma N^2\gg1$ shows
universal fluctuations.
 With increasing length $L$ of the waveguide,
the variance of the reflectance changes from
the value $2/15 N^2$, characteristic for universal conductance
fluctuations in disordered wires, to another value
$1/8 N^2$,
characteristic for chaotic cavities.
The weak-localization correction
to the average reflectance
performs a similar crossover from the value
$1/3 N$ to $1/4 N$. In the regime $\gamma N^2\ll1$,
the large-$L$ distribution of the reflectance $R$ becomes very wide
and asymmetric, $P(R)\propto (1-R)^{-2}$ for $R\ll 1-\gamma N$.
\end{abstract}

\bigskip
\narrowtext

An elegant and fundamental description of universal conductance
fluctuations is provided by random-matrix theory \cite{MPS}.
Different complex physical systems
can be classified into a few universality classes, characterized
by the dimensionality of the geometry and by the
symmetries of the scattering matrix.
The so-called circular ensemble, with a
uniform distribution of the scattering matrix
on the unitary group, describes chaotic cavities \cite{Blu89}.
The variance of the conductance in this ensemble (in units of $2e^2/h$)
equals $1/8\beta$, where $\beta=1$ ($\beta=2$) for systems with (without)
time-reversal symmetry
in the absence of spin-orbit interaction \cite{Jal94}.
This is the zero-dimensional limit, corresponding to
a logarithmic repulsion of the transmission and reflection
eigenvalues.
 A disordered wire belongs to a different universality class
(one-dimensional limit), with a non-logarithmic eigenvalue repulsion
and a variance
$2/15\beta$. The change from $1/8$ to $2/15$ is
due to a weakening of the repulsion between the smallest
transmission eigenvalues \cite{Bee93b}.

The optical analogue of universal conductance fluctuations is the
appearance of sample-to-sample fluctuations in the intensity
which is transmitted
or reflected by a random medium. Universality in this case means that
the transmitted or reflected intensity, divided by the incident intensity
per transverse mode of the medium, fluctuates with a variance which
is independent of the sample size or the degree of disorder \cite{Mel88a}.
It is essential for this universality that the incident illumination
is diffusive, which means that the incident intensity is equally
distributed over the $N$ transverse modes of the medium. A new aspect
of the optical case is the possibility of absorption or amplification
of radiation. In Ref.~\cite{Bee96} it was shown that the distribution
 of reflection
eigenvalues in the limit of an infinitely long waveguide
with absorption or amplification
is the Laguerre ensemble of random-matrix theory.
The reflection eigenvalues $R_n$, $n=1,2,\dots N$, are the eigenvalues
of the matrix product $rr^\dagger$, where $r$ is the $N\times N$ reflection
 matrix of the waveguide.  In Ref.~\cite{Bee96} the fluctuations
in the reflected intensity were computed for the case of plane-wave
illumination. The purpose of the present paper is to consider the
case of diffusive illumination, in order to make the connection with
universal conductance fluctuations.

We consider the reflection of monochromatic radiation with wavenumber
$k$ by a waveguide of length $L$ and width $W$.
The reflectance $R$ is defined as the ratio of reflected
and incident intensities. For diffusive illumination it is given by
\begin{equation}
R=N^{-1}\sum_{n,m}|r_{nm}|^2=N^{-1}\sum_n R_n.
\end{equation}
The $L$-dependence of the distribution
of reflection eigenvalues is described by the
Fokker-Planck equation\cite{Bee96}
\begin{eqnarray}
l\frac{\partial}{\partial L}P(\{\lambda_n\},L)&=&\frac{2}{\beta
N{+}2{-}\beta}\sum_{i=1}^{N}
\frac{\partial}{\partial\lambda_{i}}\lambda_{i}(1+\lambda_{i})\cr
&&\mbox{}\times
\Bigl[\frac{\partial P}{\partial\lambda_{i}}
\mbox{}+\beta P\sum_{j\neq i}\frac{1}{\lambda_{j}-\lambda_{i}}
+\gamma(\beta N{+}2{-}\beta)P\Bigr],\label{FP}
\end{eqnarray}
where we use the parametrization $R_n=\lambda_n/(1+\lambda_n)$,
$\lambda_n\in(0,\infty)$. The parameter $\gamma>0$ is the ratio of the mean
 free path $\l$ and the absorption length $l_a$.
Eq.~(\ref{FP}) is valid if $kl\gg 1$, $kl_a\gg1$, and $W\ll L$.
Optical systems normally
possess time-reversal symmetry ($\beta=1$), but in view of a recently
observed magneto-optical effect \cite{Rik96}, we also include the case of
broken time-reversal symmetry ($\beta=2$).

Relevant length scales are defined in terms of the transmitted intensity
 \cite{Dor82,Mis96}.
 The transmitted intensity decays
linearly with $L$ for $L\ll\xi$ and
exponentially for $L\gg\xi$, where the decay length
$\xi=(1/\xi_a+1/\xi_l)^{-1}$
contains a contribution $\xi_a=l(2\gamma+\gamma^2)^{-1/2}$ from absorption
and a contribution $\xi_l=\frac{1}{2}l(\beta N{+}2{-}\beta)$
 from localization by disorder.
We will study the two regimes
{\em(1)}~$\xi_a\ll\xi_l$ (or $\gamma N^2\gg1$)
and {\em(2)}~$\xi_a\gg\xi_l$ (or $\gamma N^2\ll1$). In both regimes
we assume $N\gg1$.

{\em(1)} The regime $\gamma N^2\gg1$ admits a perturbative treatment for
all $L$  by the method of moments of Mello and
Stone \cite{Mel88}. We define the moments
$M_q=N^{-1}(-1)^q\sum_i(1+\lambda_i)^{-q}$, so that $R=1+M_1$.
The Fokker-Planck equation (\ref{FP}) enables us to express derivatives
$\partial\langle M_{q_1}^{n_1}\ldots
M_{q_k}^{n_k}\rangle/ \partial L$ in terms of the moments
themselves. Expanding each moment in powers of $1/N$, we get
a closed system of first order differential equations for
the coefficients. The method has been explained in detail \cite{Mel88},
therefore we just present the result.
We need the following terms in the $1/N$-expansion of the moments,
\begin{mathletters}\begin{eqnarray}
&&\langle M_1^q\rangle=F_{q,0}+N^{-1}F_{q,1}+N^{-2}F_{q,2}+
{\cal O}(N^{-3}),\\
&&\langle M_1^qM_2\rangle=G_{q,0}+N^{-1}G_{q,1}+
{\cal O}(N^{-2}),\\
&&\langle M_1^qM_3\rangle=H_q+{\cal O}(N^{-1}).
\end{eqnarray}\end{mathletters}%
The $q$-dependence of the coefficients is given by
\begin{mathletters}\begin{eqnarray}
&&F_{q,0}=f^q,\qquad F_{q,1}=\delta_{\beta1}q\mu f^{q-1},\qquad
F_{q,2}=q\eta_\beta f^{q-1}+q(q-1)\theta_\beta f^{q-2},\\
&&G_{q,0}=gf^q,\qquad G_{q,1}=\delta_{\beta1}(\sigma f^q+q\mu g f^{q-1}),\\
&&H_q=hf^q.
\end{eqnarray}\end{mathletters}%
 Here $f$, $g$, $h$, $\mu$, $\theta_{1,2}$, $\sigma$, $\eta_{1,2}$
are functions of $L$ which obey the following system of first
order differential equations and initial conditions:
\begin{mathletters}\label{ODESystem}\begin{eqnarray}
&&ldf/dL=f^2-2\gamma f-2\gamma, \qquad f(0)=-1,\\
&&ldg/dL=4(f-\gamma)g+2(f^2-2\gamma f),\qquad g(0)=1,\\
&&ldh/dL=6(f-\gamma)h+3g(g+2f-2\gamma), \qquad h(0)=-1,\\
&&ld\mu/dL=2(f-\gamma)\mu+g-f^2, \qquad \mu(0)=0, \\
&&ld\theta_\beta/dL=4(f-\gamma)\theta_\beta+(2/\beta)(g+h)+\delta_{\beta1}\mu
(g-f^2),\qquad \theta_\beta(0)=0,\\
&&ld\sigma/dL=4(f-\gamma)\sigma+4\mu(g+f-\gamma)+2(g-f^2)+4(h-fg),\qquad
\sigma(0)=0,\\
&&ld\eta_\beta/dL=2(f-\gamma)\eta_\beta+2\theta_\beta+
\delta_{\beta1}(\sigma-2\mu f-g+f^2),\qquad \eta_\beta(0)=0.
\end{eqnarray}\end{mathletters}%
The equations can be easily solved one after another.
The first two of them determine the mean and variance of the reflectance
for plane-wave illumination, studied in Ref. \cite{Bee96}.
The other equations determine the mean and the root-mean-squared
fluctuations for diffusive illumination to order $N^{-1}$. We decompose
$\langle R\rangle=R_0+\delta R$, where $R_0={\cal O}(N^0)$
and $\delta R={\cal O}(N^{-1})$. In terms of the functions in
Eq.~(\ref{ODESystem}) we have
\begin{mathletters}\label{6}\begin{eqnarray}
&&R_0=1+f, \qquad \delta R=\delta_{\beta1}\mu/N,\\
&&\mathop{\rm{var}} R=
(2\theta_\beta-\delta_{\beta1}\mu^2)/N^2=4\theta_2/\beta N^2.
\end{eqnarray}\end{mathletters}%
The weak-localization correction $\delta R$ vanishes for $\beta=2$,
while $\mathop{\rm{var}} R\propto 1/\beta$, just as without absorption.
\begin{figure}
\hskip45mm
\epsfxsize=0.5\hsize
\epsfysize=0.5\vsize
\epsffile{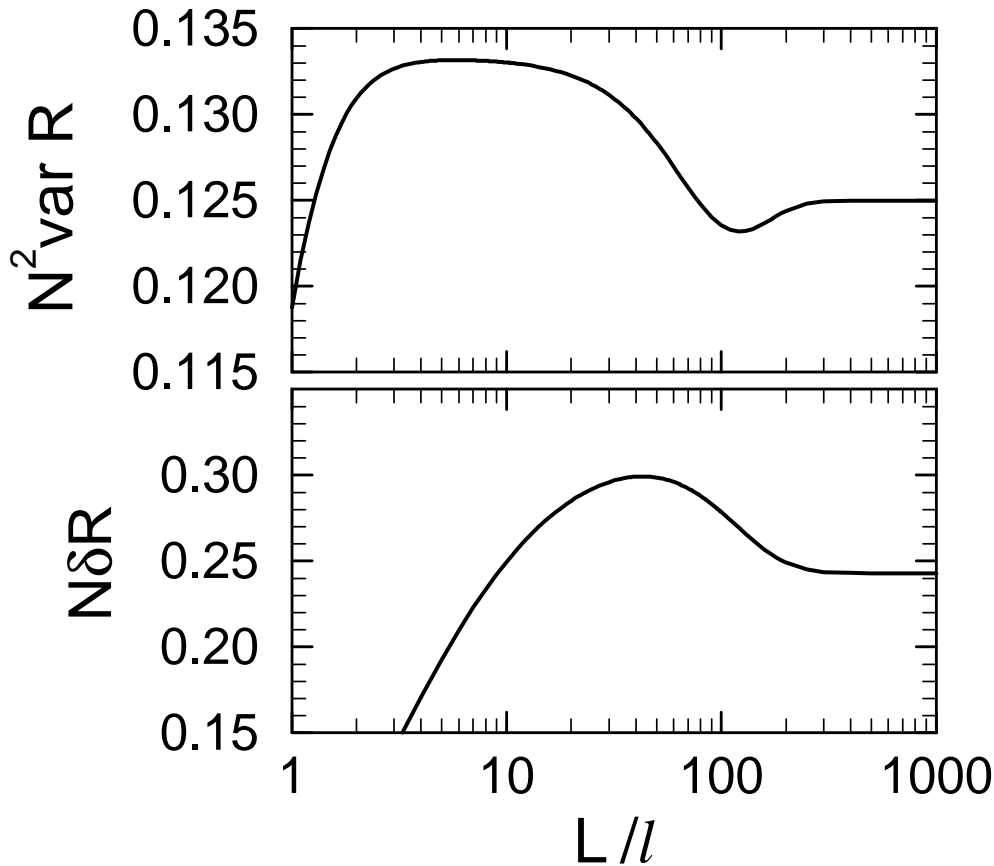}
\caption{
Length dependence of $N\delta R$ and $N^2\mathop{\rm{var}} R$
 for $\beta=1$
and $\gamma=10^{-4}$, according to
Eqs.~(\protect\ref{ODESystem}), (\protect\ref{6}), in the
regime $\gamma N^2\gg1$. The variance $\mathop{\rm{var}} R$ of the
 reflectance
crosses over from a plateau at $2/15 N^2$ (one-dimensional limit)
to a plateau at $1/8 N^2$ (zero-dimensional limit).
The crossover is non-monotonic and occurs when the length of the
waveguide becomes comparable to the decay length
$\xi=70\,l$ of the transmittance. The weak-localization correction
$\delta R$ shows a similar crossover from $1/3N$ to $1/4N$, but the
plateaus are less well-defined for this value of $\gamma$.}
\label{fig1}\end{figure}%

To find $\langle R\rangle$ and $\mathop{\rm{var}} R$ we need to solve
 the first five equations
(\ref{ODESystem}). The analytical solution is cumbersome
to use, but it is easy to integrate
the system numerically at a given $\gamma$.
Results for $\gamma=10^{-4}$ are shown in Fig.~\ref{fig1}. The large-$L$
limit can be found directly by replacing the derivatives at
the left-hand side of Eq.~(\ref{ODESystem})
by zero. In this way, we obtain the following asymptotical values:
\begin{mathletters}\label{Lag}\begin{eqnarray}
&&R_0=1+\gamma-\sqrt{2\gamma+\gamma^2},\label{RLag}\\
&&\delta R=\delta_{\beta1}N^{-1}
\left(\case{1}{2}(2+\gamma)^{-1}+\gamma-(1+\gamma)
\sqrt{\gamma(2+\gamma)^{-1}}\right),\label{WLLag}\\
&&\mathop{\rm{var}} R=\left[2\beta N^2(2+\gamma)^2\right]^{-1}.\label{VarLag}
\end{eqnarray}\end{mathletters}%
The large-$L$ limit (\ref{Lag}) can also be obtained
from the Laguerre ensemble
for the reflection eigenvalues \cite{Bee96},
\begin{equation}
P_\infty(\{\lambda_n\})\propto\exp\Bigl[\beta\sum_{i<j}\ln|\lambda_j-\lambda_i|
-\gamma(\beta N{+}2{-}\beta)\sum_i\lambda_i\Bigr],
\label{LaguerreProb}\end{equation}
which is the asymptotic solution of the Fokker-Planck equation (\ref{FP}).
The limiting values (\ref{Lag}) are reached when $L\gg\xi$, hence when
the transmittance through the waveguide has become exponentially small.
For $L\ll\xi$ the effect of absorption can be neglected. Over a range of
lengths $L$ such that $l\ll L\ll \xi$, the mean and variance
are given by \cite{Mel88} $R_0=1-l/L$, $\delta R=\delta_{\beta1}/3N$,
$\mathop{\rm{var}} R=2/15\beta N^2$.
The value $1/8\beta N^2$ for the variance when $\gamma\ll 1$ and $L\gg\xi$
follows directly from the logarithmic repulsion of the $\lambda_n$'s
in the Laguerre ensemble \cite{Bee93}.
The difference with the value $2/15\beta N^2$ for $l\ll L\ll\xi$ arises
because the repulsion is non-logarithmic in the absence of absorption
 \cite{Bee93b}.

{\em(2)} We now turn to the second regime, $\gamma N^2\ll1$.
This regime cannot be treated perturbatively for $L\gtrsim\xi$,
because of the onset of localization by disorder.
The limiting large-$L$ values of $\langle R\rangle$ and $\mathop{\rm{var}}
 R$ can be
computed from Eq.~(\ref{LaguerreProb})
using formulas \cite{Nag93} for the density and correlation function
of the $\lambda_n$'s in the Laguerre ensemble. The result is
\begin{mathletters}\label{case2}\begin{eqnarray}
\langle R\rangle&=&1-\beta\gamma N\left[\ln(1/\gamma N^2)
+{\cal O}(1)\right],\\
\mathop{\rm{var}} R&=&\beta\gamma -(\beta\gamma N)^2\left[\ln(1/\gamma N^2)
+{\cal O}(1)\right]^2.
\end{eqnarray}\end{mathletters}%
A new crossover length $L_{\rm c}=\xi\ln(1/\gamma N^2)$
appears at which the mean and variance
of $R$ attain their asymptotic values (\ref{case2}). This is the length
at which the exponential decay
$\langle 1-R\rangle\approx N^{-1}\exp(-L/4\xi)$ \cite{Zir92} comes to a halt.

\begin{figure}[ht]
\hskip45mm
\epsfxsize=0.5\hsize
\epsfysize=0.5\vsize
\epsffile{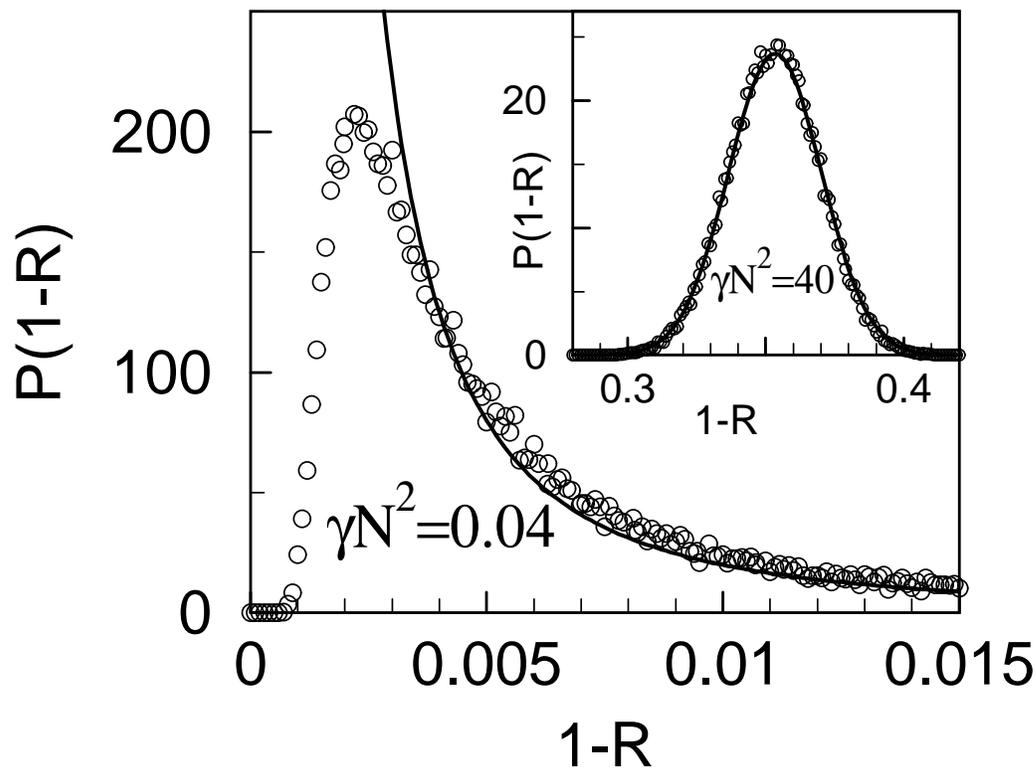}
\caption{Large-$L$ distribution of the reflectance for $\beta=1$.
Data points are
obtained by generating $5{\cdot}10^4$ random matrices in the Laguerre ensemble
(\protect\ref{LaguerreProb}). The main plot is for the regime $\gamma N^2\ll1$
($N=20$, $\gamma=10^{-4}$), the solid curve being the asymptotic tail
(\protect\ref{Rfrom1}). The inset is for the regime
$\gamma N^2\gg1$ ($N=20$, $\gamma=0.1$), the solid curve
being a Gaussian with mean and variance given by Eq.~(\protect\ref{Lag}).}
\label{fig2}\end{figure}

{}From Eq.~(\ref{case2}) we see that the reflectance has a wide distribution
in the regime $\gamma N^2\ll1$: The root-mean-squared fluctuations
of $1-R$ are greater than the mean by a factor $(\gamma N^2)^{-1/2}$.
(In the opposite regime $\gamma N^2\gg 1$, in contrast, the distribution
of the reflectance is a narrow Gaussian, see inset in Fig.~\ref{fig2}.)
To determine the
tail of the distribution, it is sufficient
to consider only the contribution from the smallest eigenvalue $\lambda_1$,
which gives the main contribution to $1-R$  when $\gamma N^2\to 0$. The
smallest eigenvalue in the Laguerre ensemble has the exponential
distribution $P(\lambda_1)=\beta\gamma N^2
\exp(-\beta\gamma N^2\lambda_1)$
\cite{Ede91}, hence
\begin{equation}
P(R)=\beta\gamma N(1-R)^{-2},\qquad \gamma N\ll 1-R\lesssim 1/N.
\label{Rfrom1}\end{equation}
We have calculated $P(R)$ by generating a large number of random
matrices in the Laguerre ensemble (see Fig.~\ref{fig2}).
The distribution reaches its maximum at $1-R\simeq\gamma N$
and then drops to zero for smaller values of $1-R$. The tail for
large values of $1-R$ is well described by Eq.~(\ref{Rfrom1}) (solid
curve in Fig.~\ref{fig2}).

To conclude, we have studied the statistics of the
reflectance in an absorbing random
waveguide under diffusive illumination.
When the decay of the transmittance is dominated by absorption
($\gamma N^2\gg1$),
the fluctuations are shown to possess the same features
as universal conductance fluctuations, including independence
on the disorder and $1/\beta$ dependence on the symmetry index.
A crossover from zero-dimensional to one-dimensional limit was found
in Refs.~\cite{Iid90,Arg96} for a chain of chaotic cavities.
(A long chain of cavities behaves as a diffusive wire.)
We have found an opposite crossover, from the
one-dimensional to the zero-dimensional limit,
as the length of the waveguide is increased beyond the decay length $\xi$.
Another regime, when the decay of the transmittance is
dominated by localization due to disorder ($\gamma N^2\ll1$),
is principally new and
characterized by a wide and asymmetric distribution of the reflectance.
The asymptotic regime establishes at a new characteristic
scale $L_{\rm c}>\xi$.

We thank P.~W.~Brouwer for helpful discussions.
This work was supported by the Nederlandse Organisatie voor
Wetenschappelijk
Onderzoek (NWO) and the Stichting voor Fundamenteel Onderzoek der
 Materie (FOM).

\end{document}